\documentclass{aa}
\usepackage{graphics}

\def\aa{\AA}
\newcommand{\teff}  {\mbox{$T_{\rm eff}$}}
\newcommand{\logg}  {\mbox{{\rm log}~$g$}}

\def\gf{{\it gf}}
\def\kms{km~s$^{-1}$}

\def\ie{i.e.}
\def\eg{e.g.}

\begin{document}

\thesaurus{08.16.3; 08.01.1; 08.01.3; 08.09.3; 02.12.3}

\title{The beryllium abundance in the very metal-poor halo star G\,64--12 
from VLT/UVES observations\thanks{Based on observations taken during the 
Science Verification of UVES at the VLT/Kueyen telescope, European Southern 
Observatory, Paranal, Chile}}

\author{
F. Primas\inst{1}\fnmsep\thanks{Visiting Researcher (CNRS) at the Laboratoire 
d'Astrophysique - Observatoire de Midi-Pyr\'en\'ees, Toulouse, France}
\and M. Asplund\inst{2}
\and P. E. Nissen\inst{3}
\and V. Hill\inst{1}
}

\offprints{F. Primas}

\institute{European Southern Observatory, Karl-Schwarzschild Str. 2, 
D-85748 Garching b. M\"{u}nchen, fprimas@eso.org, vhill@eso.org
\and Uppsala Astronomical Observatory, Box 515, SE-751 20 Uppsala, Sweden, 
martin@astro.uu.se
\and Institute of Physics and Astronomy, Aarhus University, Denmark, 
pen@ifa.au.dk}

\date{Received ; Accepted}
\authorrunning{Primas et al.}
\titlerunning{Beryllium in G\,64--12}
\maketitle

\begin{abstract}
We report on a new spectroscopic analysis of the very metal deficient 
star G\,64--12 ([Fe/H]=$-$3.3), aimed at determining, for the first time, 
its beryllium content. The spectra were observed during the Science 
Verification of UVES, the ESO VLT Ultraviolet and Visible Echelle 
Spectrograph. The high resolution ($\sim$48\,000) and high S/N ($\sim$130 
per pixel) achieved at the wavelengths of the \ion{Be}{ii} resonance 
doublet allowed an accurate determination of its abundance: log N(Be/H) 
= $-$13.10 $\pm$ 0.15~dex. The Be abundance is significantly higher 
than expected from previous measurements of Be in stars of similar 
metallicity (3D and NLTE corrections acting to make a slightly higher 
value than an LTE analysis). When compared to iron, the high [Be/Fe] ratio 
thus found may suggest a flattening in the beryllium evolutionary trend 
at the lowest metallicity end or the presence of dispersion at early 
epochs of galactic evolution. 

\end{abstract}

\section{Introduction}
The knowledge of lithium, beryllium, and boron in stars play a major role 
in our understanding of Big Bang nucleosynthesis, cosmic--ray physics, 
and stellar interiors. Lithium (together with D and He) is a key element 
to probe the Big Bang nucleosynthesis scenario, because most of $^7$Li 
is primordially produced. On the contrary, \element[][6]{Li}, 
\element[][9]{Be}, \element[][10,11]{B} (with a possible contribution to 
\element[][11]{B} from $\nu$-spallation in supernovae being still under 
debate) were suggested to originate primarily from spallation reactions in 
the interstellar medium (ISM) between cosmic-ray (CR) $\alpha$-particles and 
protons and heavy nuclei like carbon, oxygen, and nitrogen (Reeves et al. 
\cite{reeves70}). The production rate of any of these three light elements 
is then proportional to the product of the abundance of CNO (proportional 
to the number of supernovae, N$_{SN}$) and the CR flux (assumed to be 
proportional to the supernovae rate). A slope of two in the logarithmic plane 
\element[][6]{Li},\element[][9]{Be},\element[][10]{B} vs O is thus expected. 
If [O/Fe] in metal-poor halo stars is constant, then this same slope holds 
vs Fe as well. \\
On these premises, the picture emerging from the first systematic 
analyses of Be and B in stars of the Galactic halo and disk from high 
resolution data clearly challenged this scenario: both elements scale 
almost {\it linearly}, and not {\it quadratically} with metallicity from 
the early Galactic epochs up to now (\eg\ Molaro et al. \cite{molaro97}, 
Boesgaard et al. \cite{bk99b} for Be; Duncan et al. \cite{duncan97}, Primas 
et al. \cite{fp99} for B). Such finding has been interpreted as an indication 
for a primary (instead of secondary) origin of Be and B, that in turn may 
imply the need for a production mechanism independent of the metallicity 
in the ISM. Although the most recent theoretical scenarios (which seem to 
converge towards the proximity of supernovae as one of the most likely and 
efficient sites for Be and B production) are able to reproduce the observed 
abundance levels and trends, new observational inputs are still in high 
demand in order to further constrain the models. The evolution/behavior of 
Be at the lowest metallicities, the presence and magnitude of scatter (if 
any), the presence and location of a possible change of slope along the 
trend, and a successful solution of the oxygen dilemma (high or flat) in the 
Galactic halo are some of the still missing inputs. \\
With this Letter, we aim at adding a new tile to the field of 
light elements evolution, presenting the first beryllium detection in a 
star characterized by a metallicity well below one thousandth solar 
([Fe/H]=$-$3.3), the halo star G\,64--12. Our newly determined [Be/Fe] 
ratio will be compared to previous observational analyses in low metallicity 
stars and confronted with different theoretical suggestions.

\begin{table*} 
\caption{Observations log for G\,64--12}
\begin{flushleft} 
\begin{tabular}{lcccccr} \hline 
V     & Date    & Mode   & $\lambda_c$ & $\lambda$-coverage & Exp.Time & 
S/N$^{\mathrm{a}}$ \\ 
      &	        &	 &     nm      &     nm             &   min & \\ \hline
11.46 & Feb. 12 & Dic\#1 & 346 (CD\#1) & 305--385 & 2$\times$90 & 75,80 \\ 
      &         &        & 580 (CD\#3) & 480--680 & 2$\times$90 & 350,370 \\
      &	Feb. 14 & Dic\#1 & 346 (CD\#1) & 305--385 & 2$\times$80 & 65,70 \\
      &         &        & 860 (CD\#4) & 665--1050& 2$\times$80 & 270,280 \\ 
\hline
\end{tabular}
\end{flushleft}
\begin{list}{}{}
\item[$^{\mathrm{a}}$] measured on each single spectrum, around 313~nm, 510~nm 
and 700~nm in the B346, R580, and R860 settings respectively 
\end{list}
\end{table*}

\section{Observations and Data Reduction}

A total of 6 hours integration time (cf Table~1 for more details) were 
devoted to G\,64--12 (V=11.45~mag), during the Science Verification of UVES 
(February 10--18, 2000). \\
The target was observed in Dichroic mode, and the six hours were splitted 
in four exposures of $\sim$ 90~min each: two with the standard Dichroic 
\#1 setting and two with a slightly modified Dichroic \#1 setting. All the 
spectra were observed with a slit of 0.8'', corresponding to nominal 
resolving powers of 48\,000 in the blue and 55\,000 in the red, later 
confirmed by the Thorium-Argon calibration frames. The spectra were reduced 
within the MIDAS UVES context. First, bias and inter-order background were 
subtracted from both science and flat-field frames, then the object was 
optimally extracted, divided by the flat-field (extracted with the same 
weighted profile as the star), $\lambda$-calibrated and the extracted orders 
merged. All the spectra were then registered for barycentric radial velocity 
shifts, co-added, and normalized. An independent check was performed using 
standard IRAF tasks for echelle spectra reduction, which produced the same 
final product. The S/N ratio achieved at the Be~II resonance doublet on the 
final spectrum (cf Fig.~1) is 130 per pixel. 

\section{Stellar Parameters}

Using the Infrared Flux Method, Alonso et al. (\cite{alonso96}) derived 
$\teff = 6470 \pm 90$~K for G\,64--12. This is considerably higher than the 
value of \teff = 6220~K determined by Ryan et al. (\cite{ryan99}) from 
several colour indices using the ``low" \teff\ scale of Magain 
(\cite{magain87}). It should be noted, however, that Ryan et al. adopt a 
negligible low interstellar reddening of G\,64--12, whereas the Str\"omgren
$uvby$-$\beta$ photometry as given in their Table 1 suggests a reddening
$E(b-y) = 0.023$ corresponding to $E(B-V) \simeq 0.03$. Correcting the colour
indices for this reddening would increase the Ryan et al. temperature to
about 6360~K. We conclude that the effective temperature of G\,64--12
is quite uncertain; depending on the adopted reddening and the zero-point 
of the \teff\ scale, one derives values between 6200 and 6500 K. However, 
fitting of the H$\alpha$ wings indicates a temperature closer to the 
higher end of the considered range of values (\teff$\simeq$ 6400~K). \\
The Hipparcos parallax of G\,64--12 ($\pi = 1.88 \pm 2.90$~mas) is too
uncertain to be used to derive the surface gravity of the star. We note,
however, that the position of G\,64--12 in the $(b-y)_0 - c_0$ diagram
(see Ryan et al. \cite{ryan99}, Fig. 3) is similar to that of HD~84937. 
Using the absolute magnitude calibration of Nissen \& Schuster 
(\cite{nissen91}), $M_V \simeq 3.6$ is derived for both stars. Hence, the 
gravity of G\,64--12 must be close to that of HD~84937 (\logg = 4.1) as 
derived from its Hipparcos parallax (Nissen et al. \cite{nissen97}). 
Altogether, we estimate that the gravity of G\,64--12 lies in the range 
\logg = 3.9 - 4.3. \\
The metallicity of G\,64--12 was derived from equivalent widths of 22
\ion{Fe}{i} lines in the spectral region 480 - 560~nm. The lines are weak 
($2 < EW < 30$~m\AA ), but the EW's are very reliable due to the high S/N. 
Adopting $gf$-values from O'Brian et al. (\cite{obrian91}) and using a 1D 
model atmosphere with \teff\ = 6400, [Fe/H] = $-3.28 \pm 0.08$~dex is 
derived. Allowing for the possible range of \teff\ (6300 - 6500~K),
[Fe/H] = $ -3.28 \pm 0.10$. \\
It is worth noticing that the same procedure applied to 5 \ion{Fe}{ii} 
lines yields a somewhat lower iron content [Fe/H] = $ -3.42~\pm~0.10$. 
Though this would be in principle our preferred metallicity indicator 
(\ion{Fe}{ii} lines are much less sensitive to NLTE corrections), the 
\gf-values of these lines are not very accurately known. Four out of five 
are listed in the compilation by Kroll \& Kock (\cite{kk87}), and all 
five are found in the work by Moity (\cite{moity83}). Unfortunately, there 
is a systematic difference of $\sim$0.20~dex between these two investigations 
(Moity's values being lower). This has some consequences on the final iron 
abundance derived from these (few) lines, and certainly has a non-negligible 
effect on the derivation of \logg~from the ionization balance. We ran a few 
tests and reached the ionization balance for \logg=4.3 if mean {\it gf}-values 
were adopted, and for \logg=4.1 if Moity's values were used. These results, 
together with the paucity of \ion{Fe}{ii} lines, made us decide to keep 
\logg = 4.1 $\pm$ 0.2 as our final choice for the analysis.

\begin{figure}[t*] 
\vspace{-0.3cm}
\resizebox{9cm}{9cm}{\includegraphics{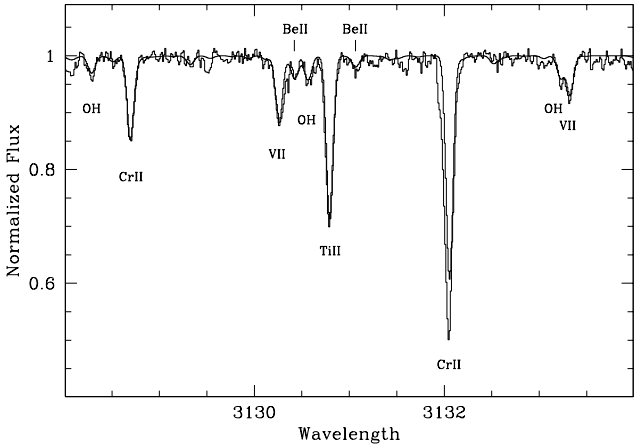}}
\vspace{-3cm} 
\caption{Reduced and normalized observed spectrum of G\,64--12 
with overplotted our best-fit synthesis (thin line)}
\label{fig1} 
\end{figure}
\section{Abundance Analysis Highlights}

The analysis of the beryllium spectral region was performed via spectrum 
synthesis. For this purpose, we adopted a model atmosphere computed with 
\teff~= 6400~K, \logg = 4.1, and [Fe/H] = $-$3.3. Microturbulence was 
assumed to be 1.5~\kms. \\
The 1D spectrum synthesis analysis was carried out with Kurucz model 
atmospheres and SYNTHE (Kurucz \cite{kur93}). The model was convolved with 
a Gaussian function of FWHM = 6.7~\kms~(which combines instrument, 
rotation and macroturbulence broadenings), and was computed with the 
$\alpha$-elements enhanced by +0.4~dex and with the ``approximate 
overshooting'' option switched off. The solar abundances were taken from the 
compilation of Grevesse \& Sauval (\cite{gs98}). The list of lines necessary 
to synthesize the near-UV spectral region is the one tested by Primas et al. 
(\cite{fp97}). \\
Additionally a time-dependent 3D hydrodynamical model atmosphere
with parameters appropriate for G\,64--12 has been constructed, with the 
same 3D, compressible, radiative-hydrodynamics code which has previously been 
successfully used for studies of solar and stellar convection 
(e.g. Stein \& Nordlund \cite{sn98}; Asplund et al. \cite{asp99}, 
\cite{asp00}). The equations of mass, momentum and energy coupled to the 
3D radiative transfer equation (which includes the effects of 
line-blanketing) have been solved on a Eulerian mesh with 
$100\times 100\times 82$ zones covering $21\times 21\times 8.5$~Mm. For the 
3D LTE spectrum synthesis, a representative sequence of 1\,hr stellar time 
with snapshots every 1\,min was selected; the temporal average of \teff\ 
for the sequence is $6460\pm27$\,K, i.e. very close to the value indicated 
by the IRFM method. It is noteworthy that no free parameters like mixing 
length parameters, micro- and macroturbulence enter the construction of 3D 
model atmospheres or spectral synthesis, as the time-dependent convective 
motions and inhomogeneities are self-consistently calculated. In order to 
quantify the systematic errors introduced when relying on 1D analyses, a 
strictly differential comparison of the 3D predictions have been carried 
out with a corresponding MARCS model (Gustafsson et al. \cite{gustf75}) with 
the same parameters and input data (Asplund et al. \cite{asp97}).

\subsection{On other Elemental Abundances}
Because of the large spectral coverage (see Table 1), all the most commonly 
studied elements ($\alpha$, iron-group, heavy) are accessible from the 
observed spectrum of G\,64--12. Although the complete abundance analysis of 
G\,64--12 is beyond the scope of this work, some of the elements most relevant 
to it deserve mentioning. \\
\noindent{\it $\alpha$-elements}: we checked the abundance of magnesium and 
calcium in order to test how representative G\,64--12 is of the population of 
the very metal-deficient stars. From the equivalent widths measured for two 
\ion{Mg}{i}b lines and four \ion{Ca}{i} lines we derive [Mg/Fe] = +0.43 $\pm$ 
0.03~dex~and [Ca/Fe] = +0.48 $\pm$ 0.07~dex, \i.e.\ ``normal'' and in good 
agreement with what can be found in the literature. \\
\noindent{\it Oxygen (from OH lines)}: the same spectral region where the 
\ion{Be}{ii} doublet falls is also very rich in OH molecular lines, which 
represent one of the few oxygen abundance indicators available to 
spectroscopists. Our 1D LTE spectrum synthesis analysis suggests [O/Fe] = 
+1.1, similarly to what found by Israelian et al. (\cite{isr98}) and 
Boesgaard et al. (\cite{bk99a}) using the same abundance indicator and for 
stars of comparable metallicity. Our finding is also in agreement with the 
recent oxygen determination made by Israelian et al. (\cite{isr2000}) in the 
same star. However, due to severe systematic errors affecting the near-UV OH 
lines in metal-poor stars, we prefer not to adopt this value. We suspect 
that the oxygen abundance derived from the UV OH lines is in error, being 
affected by metallicity dependent 3D and/or NLTE effects (Asplund et al. 
\cite{asp99}). A future analysis of Be in a larger sample of stars will deal 
with this issue (Primas et al., in preparation). \\
\noindent{\it Lithium}: the S/N ratio achieved around the \ion{Li}{i} line 
at 670.8~nm is $\approx$ 300. Our measurement of the \ion{Li}{i} 670.8nm 
equivalent width gives EW = 22.2 $\pm$ 0.5~m\aa, in agreement with the value 
published by Ryan et al. (\cite{ryan99}, EW = 21.2 $\pm$ 1.1~m\aa). The 
lithium abundance derived is A(Li) = 2.23 $\pm$ 0.01~dex. The difference 
with Ryan et al. (\cite{ryan99}) value (A(Li) = 2.14 $\pm$ 0.03~dex) can be 
explained by the difference in adopted temperatures. 

\section{The Beryllium Abundance and Its Uncertainty}

The best-fit synthesis of the \ion{Be}{ii} doublet is achieved with 
$\log$ N(Be/H)= $-13.20$ for the 313.0~nm line, and with $\log$ N(Be/H)= 
$-13.10$ for the 313.1~nm line. Being both lines well resolved, equal 
weights were assigned. The beryllium abundance thus derived using Kurucz 1D 
LTE model atmosphere is $\log$ N(Be/H)= $-13.15 \pm 0.15$~dex, significantly 
higher than what is expected from extrapolating the previously observed 
trends to these low metallicities (under the assumption that the Be abundance 
keeps decreasing). The magnitude of the error-bar assigned to our measurement 
mainly comes from the dependence of the beryllium abundance on changes in 
the stellar parameters. In order of importance, we find that $\Delta$Be = 
$\pm$0.1 dex for $\Delta\logg = \pm~0.2$, $\Delta$Be = $\pm$0.04 dex for 
$\Delta\teff = \pm~100~K$, and $\Delta$Be = $\pm$ 0.03 dex for 
$\Delta$[Fe/H] = $\pm~0.1$ dex. To these values (to be summed in quadrature) 
one has to add the dependence of Be on the placement of the continuum. This 
uncertainty in such metal-poor star is estimated to be $\pm$1\% at most, 
implying a variation in Be of $\pm$0.05~dex. Fig.~2 shows 
an enlarged view of the Be doublet, with our 1D LTE best-fit synthesis, 
and two others computed with $\pm$~0.15~dex variation in the Be content, 
which demonstrate our detection and measurement accuracy. 

\begin{figure}
\vspace{-0.3cm} 
\resizebox{9cm}{9cm}{\includegraphics{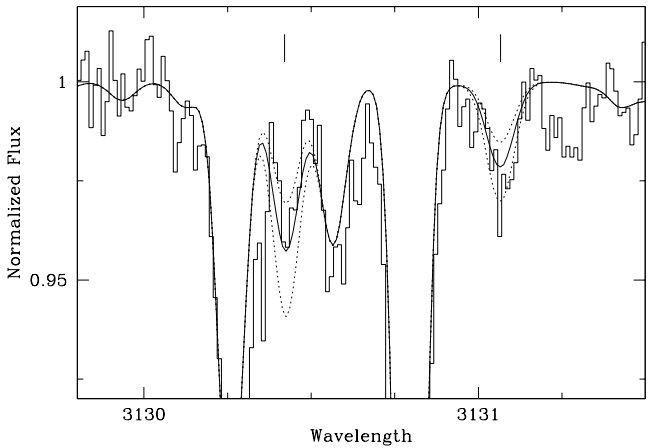}}
\vspace{-3cm} 
\caption{An enlarged view of the \ion{Be}{ii} resonance doublet in G\,64--12 
(histogram), together with our best-fit synthesis (thin line). Overplotted 
are two syntheses computed with the beryllium abundance increased 
and reduced respectively by 0.15~dex (dotted lines)}
\label{fig2} 
\end{figure}

As expected, the predicted \ion{Be}{ii} line strengths with the 3D model 
atmosphere are quite similar to the 1D case, because the lines are formed in 
the deep atmospheric layers due to the lines being very weak and coming 
from the majority ionized species. Assuming LTE, the abundance correction 
derived by using 3D hydrodynamical model atmospheres is only ~0.03~dex for 
\ion{Be}{ii} 313.1 in G\,64--12, in the sense that the 1D abundance should 
be corrected upwards with 0.03~dex. It should be emphasized that the 1D-3D 
comparison has been done strictly differentially. The quoted differences are 
therefore estimates of the systematic errors due to the adopted model 
atmospheres one can expect when using a 1D analysis, regardless of whether 
Kurucz or MARCS model atmospheres have been used. \\
Although the above-mentioned 3D effects are small, it should be borne in 
mind that these estimates assume LTE and there is no guarantee that the 
same will be true for NLTE. Recent improved 1D NLTE calculations 
appropriately done for G\,64-12 with a very extended Be atom (71 levels 
compared to the 9 used by Garc{\`\i}a L{\`o}pez et al. \cite{garcia95}), 
find some small but significant NLTE effects. In terms of abundance, the 
NLTE abundance is 0.07~dex larger than the LTE result. In spite of the 
similarity with the LTE result, it should be noted that the line formation 
for the \ion{Be}{ii} lines is far from being in LTE, since the dominant 
NLTE effects -- over-ionization and over-excitation -- have opposite 
influence on the line strength and thus the effects partly cancel. This may 
suggest that for some other stellar parameters the NLTE effects could be 
significantly larger. The details of the 1D NLTE calculations for 
\ion{Be}{ii} appropriate for a range of stellar parameters will be presented 
elsewhere (Garc{\`\i}a Perez et al., in preparation). It also remains to be 
investigated whether the steep temperature gradients and inhomogeneities in 
3D model atmospheres may drive more pronounced departures from LTE, similar 
to recent calculations for Li and O (Asplund \& Carlsson \cite{aspc00}; 
Asplund et al., in preparation). 

\section{The [Be/Fe] ratio}

Our claim to have detected a significantly higher beryllium abundance in 
G\,64--12 (compared to other low metallicity stars) can be more clearly 
seen when our new (1D LTE, for consistency with the rest of the sample) 
determination is plotted in a $\log$ N(Be/H) vs. [Fe/H] graph (cf Fig.~3a) 
together with carefully selected literature data of similar quality. 
All the data shown in Fig.~3 (both panels) had their Be abundances computed 
assuming a ``high'' \teff~scale, hence consistent with what we did. 

\begin{figure}[t]
\vspace{-0.5cm}
\hspace*{-0.8cm}\resizebox{9.6cm}{13.4cm}{\includegraphics{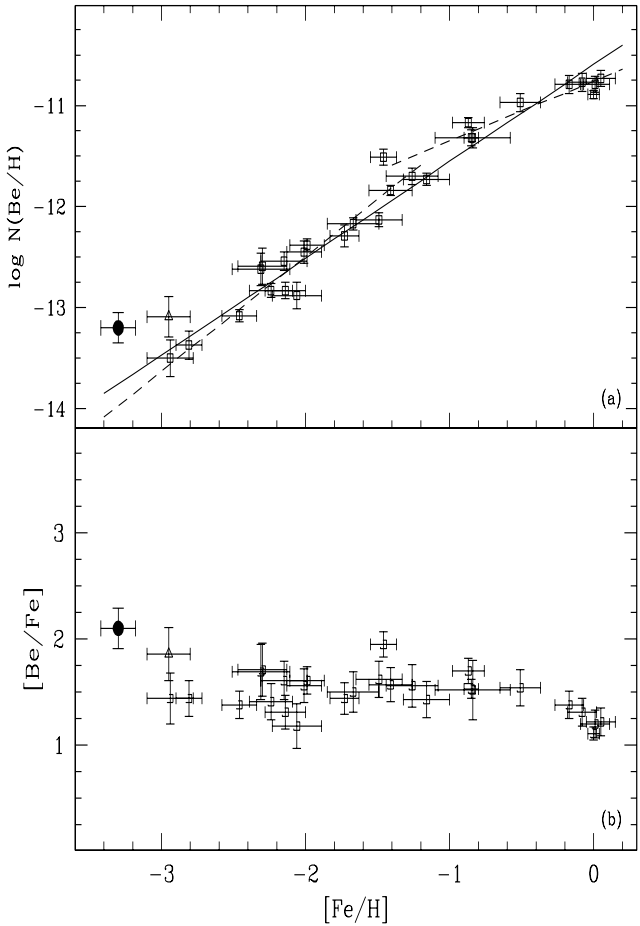}}
\caption{Panel (a): $\log$ N(Be/H) vs. [Fe/H] for the highest quality data 
points currently available, with the Boesgaard et al. (\cite{bk99b}) least 
square fits overplotted (open squares from Boesgaard et al. (\cite{bk99b}), 
the open triangle is LP 815--43 from Primas et al. (\cite{fp00}), whereas 
the filled symbol represents this work -- our 1D LTE value, for consistency 
with the rest of the data points); Panel (b): [Be/Fe] vs [Fe/H] for all the 
stars plotted in Panel (a) (same symbols have been adopted)}
\label{fig3} 
\end{figure}

Extrapolating the Boesgaard et al. (\cite{bk99b}) Be vs Fe relation at 
lower metallicities, our 1D LTE Be value is 0.6~dex higher (using their 
single line fit) or 0.8~dex higher (using their double line fit), regardless 
of the \teff~scale adopted. One may also compare G\,64--12 to BD$-$04
\degr3208 (similar \teff~ and \logg~ but a factor of 10 higher metallicity, 
cf Boesgaard et al. \cite{bk99b}) and note that their Be contents differ 
by a factor of 3 only (instead of the expected factor of 10, should Be and 
Fe grow at the same rate). From our measurements we also infer [Be/Fe] = 
2.15 $\pm$ 0.19, which is at least 0.5~dex off the average trend (Fig.~3b). 
Although we must say that G\,64--12, together with the recent Be detection 
in LP 815--43 ([Fe/H] = $-$2.95, open triangle in Fig.~3; cf Primas et al. 
\cite{fp00}), gives more weight to the hypothesis of a possible flattening 
of the Be trend at the lowest metallicities, at the moment, the interpretation 
of our new result is not straightforward.  \\
According to Vangioni-Flam et al. (\cite{vf98}), the finding of high Be 
abundances in stars with metallicity below [Fe/H] = $-$3.0 may favor shock 
acceleration in the gaseous phase of superbubbles produced by collective 
SNII explosions as the main mechanism of Be production. This, in turn, 
implies that the major role is played by the most massive stars only 
(say, with initial mass M $\geq$ 60M$_{\sun}$). \\
According to the Parizot \& Drury (\cite{pd00}) scenario, a double trend 
of Be vs O (and Fe) is expected depending on if the star formed 
inside a superbubble (hence with a high Be content due to the effect of 
collective SN explosions) or outside. In this case, we cannot exclude the 
possibility that G\,64--12 may be representative of the ``high'' 
trend, whereas other previously studied very metal-deficient stars 
(like BD$-$13\degr3442, for instance, Boesgaard et al. \cite{bk99b}) formed 
out of gas outside a superbubble. In order to test this hypothesis, new 
detections of beryllium in this challenging metallicity range are needed. \\
One last possibility that should not be underestimated is the effect due 
to 3D and NLTE corrections. As previously stated, NLTE corrections have 
usually been assumed to be negligible, but they could be important for 
different combinations of stellar parameters. The future results of 
Garc{\`\i}a Perez et al. work will shed light on this issue.  

\section{Conclusions}
We have presented a new analysis of the near-UV part of the spectrum of 
G\,64--12 with the aim of studying its beryllium content. The high 
quality of this UVES (Science Verification) spectrum has led to a clear 
detection of beryllium. This is the first measurement of Be in a star of 
such low metal content ([Fe/H] = $-$3.3). We have performed our analysis 
with 1D LTE model atmospheres, but introduced the novelty of correcting it 
for both 3D and non-LTE effects. Hence, our final 3D NLTE Be abundance is 
$\log$ N(Be/H) = $-13.05 \pm 0.15$. We have shown that when compared in a 
consistent manner to previous high quality literature data, the Be abundance 
here detected is significantly higher than what expected. The strength of 
the OH molecular lines is comparable to what observed in other stars of 
similar metallicity, thus excluding the possibility that the high Be content 
measured in G\,64--12 may be somehow related to an abnormally high O content. 
Furthermore, our measurement disagrees with the statement of a parallel (\ie\ 
same rate) increase of Be and Fe, which, in turn, suggests a flattening or 
the presence of dispersion at given metallicity in the Be vs Fe trend during 
the early evolutionary phases of our Galaxy. 

\begin{acknowledgements}
We thank the Science Operations (on Paranal) and the UVES Science 
Verification Teams for the efforts devoted to taking the observations and 
releasing the data to the ESO member states. A. Garc{\`\i}a Perez is thanked 
for making available to us her first results on new NLTE Be calculations, 
as well as E. Depagne for checking the H$\alpha$ fitting procedure. An 
anonymous referee is thanked for the quick report and the positive comments.  
\end{acknowledgements}


\end{document}